\documentclass[12pt,a4paper]{article}
\usepackage[dvips]{graphicx}
\usepackage{amssymb}
\usepackage{latexsym}

\makeatletter 

\@addtoreset{equation}{section}
\makeatother

\def\rem#1{}

\def\gl{{\mathfrak{gl}}}
\def\C{{\mathbb{C}}} % \C=\mathbb{C}
\def\R{{\mathbb{R}}} % \R=\mathbb{R}
\def\I{{\mathbb{I}}} % \Z=\mathbb{I}
\begin{document}

\setlength{\oddsidemargin}{0cm}
\setlength{\baselineskip}{6.3mm}

\begin{titlepage}
    \begin{normalsize}
     \begin{flushright}
            KEK-TH 1433 \\
            % {preliminary version} \\
            % 
     \end{flushright}
    \end{normalsize}
    \begin{LARGE}
       \vspace{1cm}
       \begin{center}
        {Localization of Vortex Partition Functions in $\mathcal{N}=(2,2) $ Super Yang-Mills theory  %\\
                  }
       \end{center}
    \end{LARGE}

   \vspace{5mm}

\begin{center}
          Yutaka Yoshida \\
        % \footnote{E-mail address: yyoshida@post.kek.jp} \\
       \vspace{4mm}
                  {\it KEK Theory Center, Institute of Particle and Nuclear Studies,} \\
                  {\it High Energy Accelerator Research Organization                   } \\
                  {\it Tsukuba, Ibaraki 305-0801, Japan                   } \\
       \vspace{0.5cm}
                  {E-mail address: yyoshida@post.kek.jp} \\
        \vspace{2cm}
 \begin{abstract}
In this article, we study the localizaiton of  the partition function of BPS vortices in 
$\mathcal{N}=(2,2)$ $U(N)$ super Yang-Mills theory with $N$-flavor  on $\R^2$.  
The vortex partition function  for $\mathcal{N}=(2,2)$ super Yang-Mills theory is obtained  from
the one in $\mathcal{N}=(4,4)$ super Yang-Mills theory by mass deformation. We show  that the partition function can be written as $Q$-exact form 
and integration in  the partition functions is localized to the fixed points  which  are related to  $N$-tuple  
one dimensional partitions of positive integers.
\end{abstract}

\end{center}
\end{titlepage}
%\end{titlepage}
\vfill

\date{\today}
%\end{frontmatter}}

\section{Introduction}

The vortices in  two dimensional
field theory are instanton like objects and have finite energy and action. We can consider the partition functions of vortices.
In supersymmetric theories with eight  or four  supercharges,  our insterest is the  vortices described by Bogomoln'yi equation
which preserve  half of the supersymmetries.
 The vortex moduli spaces in $p$-dimesional theories with eight supercharges
 are constructed in D($p$-2)-D$p$ system in \cite{HT1},  as the  K$\ddot{\mathrm{a}}$hler quotient spaces.  
Especially,  the vortex partition functions for $\mathcal{N}=(4,4)$ super Yang-Mills theories with vortex number $k$ are  $U(k)$ gauged matrix model with four supercharges \cite{HT1}, \cite{CT}. But,  it is difficult to   perform  multi-variable integrations directly in vortex partition functions.  

Another road  to evaluate the vortex partitions is the reduction from  instanton partition functions in four dimensional $\mathcal{N}=2$ super Yang-Mills theory with  surface operators. The instanton partition functions with surface operators are expanded   in double series with respect to vortex number and instanton number.
When   surface operators are  half-BPS in $U(N)$ $\mathcal{N}=2$ super Yang-Mills theory, the theory on the plane  where the surface operators  define  becomes $\mathcal{N}=(2,2)$ $U(1)$ gauge theory with  matter fields.
The insertion of surface operators in four dimensional gauge theory  correspond to  introduction of open string in topological A-model amplitudes in toric Calabi-Yau 3-folds \cite{KPW} which can be calculated by refined topological vertex \cite{IKV}. 
The authors of \cite{DGH} calculated abelian vortex partition functions in $\mathcal{N}=(2,2)$ SQED by taking decoupling limit of
instaton parts.

In this article, we consider the evaluation of the vortex partition functions for $\mathcal{N}=(2,2)$ $U(N)$ gauge theory in two dimensions with $N_f=N$ fundamental chiral multiplets  and with generic twisted masses.
This article is organized as follow: In section two, we first explain  the Bogomoln'yi equation in 2d $\mathcal{N}=(4,4)$ supersymmetric gauge theory and present vortex partition functions. In section three, deforming  the $\mathcal{N}=(4,4)$ theory by adding a superpotential term
and taking large mass limit, we obtain the vortex partition functions in $\mathcal{N}=(2,2)$ theory. In section four, we evaluate
the  vortex partitions by equivariant localization method. In section five, we also calculate K-theoretic vortex partition function from the equivariant character  of the K$\ddot{\mathrm{a}}$hler quotient space. In section six, we discuss  our results.

%%%%%%%%%%%%%%%%%%%%%%%%%%%%%%%%%%%%%%%%%%%%%%%%%%%%%%%%%
\section{Vortex in $\mathcal{N}=(4,4)$ $U(N)$ super Yang-Mills theories}
In this section, we review $\mathcal{N}=(4,4)$ $U(N_c)$ super Yang-Mills theory in two dimensions \cite{Tong1}
 with the number of hypermultiplets $N_f=N_c=N$. 
%We often write $N_c$ and $N_f$ explicitely.
 We can construct $\mathcal{N}=(4,4)$ multiplets by combining a pair of  $\mathcal{N}=(2,2)$ multiplets.
The vector multiplet in $\mathcal{N}=(4,4)$ consists of a pair of  $\mathcal{N}=(2,2)$ superfields $(\Sigma, \Phi)$.
\begin{eqnarray}
&&\Sigma=\sigma +i\sqrt{2} \theta^+ \bar{\lambda}_+ -i\sqrt{2} \bar{\theta}^- {\lambda}_{-} +\sqrt{2} \theta^+ \bar{\theta}^-
(D^3-i F_{12}) + \cdots, \nonumber \\
&&\Phi=\tilde{\sigma} +\sqrt{2} \theta^+ \tilde{\lambda}_+ + \sqrt{2} \theta^- \tilde{\lambda}_{-} +\frac{1}{\sqrt{2}} \theta^+ \theta^-
(D^1-i D^2) + \cdots, 
\end{eqnarray}
where $\Sigma$ is the twisted chiral multiplet   and  $\Phi$ is the chiral multiplet in adjoint representation in $U(N_c)$.
Here $\sigma $ and $\tilde{\sigma}$ are scalars, $\lambda $ and $\tilde{\lambda}$ are fermions.
$D^i  (i=1, 2, 3)$ are auxiliary fields.  
The hypermultiplet in $\mathcal{N}=(4,4)$ consists of  a pair of  $\mathcal{N}=(2,2)$ chiral  superfields  $(Q_i, \tilde{Q}_i), (i=1, \cdots, N_f)$.
\begin{eqnarray}
Q_i&=&q_i +\sqrt{2} \theta^+ {\psi}_{+ i}  +\sqrt{2} \theta^- {\psi}_{- i} + \theta^+ \theta^-
F_i + \cdots, \nonumber \\
\tilde{Q}_i&=&\tilde{q}_i  + \sqrt{2} \theta^+ \tilde{\psi}_{+ i}  +\sqrt{2} \theta^- \tilde{\psi}_{- i} + \theta^+ \theta^-
\tilde{F}_i + \cdots, 
\end{eqnarray}
where $Q$ are chiral multiplets in fundamental representation for $U(N_c)$  and $N_f$ flavor.
$\tilde{Q}$ are chiral multiplets in anti-fundamental representation for $U(N_c)$  and $N_f$ flavor.
 
The Lagrangian for 2d $\mathcal{N}=(4,4)$ super Yang-Mills theory with gauge group $U(N)$ and $N$ flavor consists of 
\begin{eqnarray}\label{Lag44}
S_K&=& \int d^2 x  d^4 \theta (\mathrm{Tr} \bar{\Sigma} \Sigma + \mathrm{Tr} \bar{\Phi} e^{V} \Phi e^{-V} +{Q}^{\dagger}_i e^{-2V} {Q}_i
+\tilde{Q}^{\dagger}_i e^{2V} \tilde{Q}_i  ),  \nonumber \\
S_V&=&\int d^2 x d^2 \theta \tilde{Q}_i \Phi {Q }_i.
\end{eqnarray}
We can also include  the Fayet-Iliopoulos terms and the  bosonic part in the Lagrangian becomes 
\begin{eqnarray}\label{Lagbos44}
L_{boson}&=&- \Bigl[ \mathrm{Tr} \Bigl( \frac{1}{2e^2} F_{\mu \nu} F^{\mu \nu} +\frac{1}{2 e^2} D_{\mu} \sigma D^{\mu} \bar{\sigma}
+\frac{1}{2 e^2} D_{\mu} \tilde{\sigma} D^{\mu} \bar{\tilde{\sigma}} 
+\frac{e^2}{2} ( {q}_i  {q}^{\dagger}_i -\tilde{q}^{\dagger}_i  \tilde{q}_i  -r \mathbb{I}_N)^2 \Bigr) \nonumber \\
&&+
D_{\mu} {q}^{\dagger}_i D^{\mu} q_i +D_{\mu} \tilde{q}^{\dagger}_i D^{\mu} \tilde{q}_i 
+\tilde{q}^{\dagger}_i \{\bar{\sigma}  , \sigma\}  \tilde{q}_i  + \cdots \Bigr].
\end{eqnarray}
%we can set $r^1+ir^2=0$ by $SU(2)$ R-symmery.
When the Fayet-Iliopoulos parameter $r$ is positive, the vacuum is unique up to Weyl symmetry,  
\begin{eqnarray}
q^{a}_{i}=\sqrt{r} \delta^{a}_{i}, \quad \tilde{q}^{a}_{i}=0, \quad \sigma=0, \quad \tilde{\sigma}=0,
\end{eqnarray}
where we write color indices $a (a=1, \cdots, N)$ and flavor indices $i (i=1, \cdots, N)$ explicitly.
In this phase, gauge and flavor symmetry group  breaks to  little symmetry group
\begin{eqnarray}
U(N)_{G} \times SU(N)_{F}  \to SU(N)_{\mathrm{diag}}.
\end{eqnarray}
In this case, half-BPS equation exist which minimize the energy.
This  half-BPS  solutions of  equation(Bogomoln'yi equation) are defined by
\begin{eqnarray}
&&{F_{12}}^{a}_{b}=\frac{e^2}{2} (q^{a}_{i} {q^{i}_{b}}^{\dagger}- r \delta^{a}_{b}), \nonumber \\
&&({D_{z}} q)^{a}_{i}=0,
\end{eqnarray}
where we define covariant derivative $D_z=D_{1}-i D_2$.

The vortex instanton number is given by the first Chern number $k$; 
\begin{eqnarray}
k:=\frac{1}{2\pi}\int \mathrm{Tr} F. 
\end{eqnarray}
The bosonic solution with  vortex number-$k$ has action
\begin{eqnarray}
S_{k}=2 \pi (r+ i\theta) k. 
\end{eqnarray}

Next we consider the $k$-vortex partition function for $\mathcal{N}=(4,4)$ super Yang-Mills theories.
%The moduli  of Bogomoln'yi equation of $\mathcal{N}=4$ SQCD in three dimensions was first considered by D1-D3 system and
%they consist of $\mathcal{N}=(2,2)$ super symmetric quantum mechanics.
In two dimensions,  the vortex partition function in $\mathcal{N}=(4,4)$ Yang-Mills theories is constructed by  $k$ D0-branes and $N$ D2-branes system.
This is  called the vortex matrix model \cite{CT}.

This  model is obtained by  dimensional reduction of $\mathcal{N}=1$ supersymmetric action in four dimensions to
zero dimension or equivalently dimensional reduction of $\mathcal{N}=(2,2)$ supersymmetric theory in two dimensions to zero dimension. 

This matrix model is given by
\begin{eqnarray}\label{knvortex}
Z^{\mathcal{N}=(4,4)}_{k, N} &=&\frac{1}{\mathrm{Vol} (U(k))} \int \mathcal{D} X   \exp (-\frac{1}{g^2} S_G-S_{m}-S_{A}),
\end{eqnarray}
where
\begin{eqnarray}\label{vortexmatrix}
S_G&=&\mathrm{Tr} \Bigl( -\frac{1}{2}[\varphi, \bar{\varphi}]^2-\frac{1}{2}[\phi, \bar{\phi}]^2+[\bar{\varphi}, \bar{\phi}]
[{\varphi}, {\phi}] +[{\varphi}, \bar{\phi}][\bar{\varphi}, {\phi}] + D^2 +g^2 \zeta D \nonumber \\
&&+\sqrt{2} (\bar{\lambda}_{+}  [\bar{\varphi}, \lambda_{+}] +\bar{\lambda}_{-}  [{\varphi}, \lambda_{-}]
- \bar{\lambda}_{+}  [\bar{\phi}, \lambda_{-}] -\bar{\lambda}_{-}  [{\phi}, \lambda_{+}] )
  \Bigr), \nonumber \\
S_{m}&=&\sum_{i=1}^{N} \Bigl( -I^{\dagger}_{ i} \{ \varphi, \bar{\varphi} \} I_{i}  - I^{\dagger}_{ i} \{ \phi, \bar{\phi} \} I_{i} 
 +I^{\dagger} DI \nonumber \\
&&+\sqrt{2} ({\mu}^{\dagger}_{+ i} \bar{\varphi}  \mu_{+ i} + {\mu}^{\dagger}_{- i} {\varphi}  \mu_{- i}
- {\mu}^{\dagger}_{+ i} \bar{\phi}  \mu_{- i } - {\mu}^{\dagger}_{- i} {\phi}  \mu_{+ i} )\nonumber \\ 
&&+i \sqrt{2} [ I^{\dagger i} (\lambda_{-} \mu_{+ i} -\lambda_{+} \mu_{- i} ) 
+ (-{\mu}^{\dagger}_{+ i} \bar{\lambda}_{-}  -{\mu}^{\dagger}_{- i} \bar{\lambda}_{+ }  ) I_i ] \Bigr), \nonumber \\
S_A&=&\mathrm{Tr} \Bigl( | [\varphi, B^{\dagger }] |^2 +| [\phi, B^{\dagger }] |^2 +D[B, B^{\dagger}]  \nonumber \\
&&\sqrt{2} (-[\varphi, {\rho}^{\dagger}_- ]   \rho_{-} -[\bar{\varphi},{\rho}^{\dagger}_+ ]   \rho_{+})
-[\phi,{\rho}^{\dagger}_+ ]   \rho_{+} -[\bar{\phi}, {\rho}^{\dagger}_+ ]   \rho_{+} 
\nonumber \\
&&+  i \sqrt{2} \Bigl[ B([ {\rho}^{\dagger}_{-} , \bar{\lambda}_{+} ]+[ {\rho}^{\dagger}_{+} , \bar{\lambda}_{-} ])+ B^{\dagger } ([ \lambda_{+},  \rho_{ -} ] +[ \lambda_{-},  \rho_{ +} ] ) \Bigr] \Bigr). 
\end{eqnarray}
$(\varphi, \bar{\varphi}, \phi, \bar{\phi}, \lambda, \bar{\lambda}, D) $ come from the 2d  $\mathcal{N}=(2, 2)$ vector multiplet in the 
reduction.
$(I_{i}, \mu_{  \pm i}), (i=1, \cdots, N)$ appear through dimensional reduction of the 2d  chiral multiplets  to zero dimension, which belong to fundamental representation of $U(k)$.
$(B, \rho_{\pm})$ is the dimensional reduction of 2d chiral multiplet to zero dimension, 
 adjoint representation of $U(k)$.
$\mathcal{D} X$ is the integration measure. 
The Fayet-Illioporos parameter in the vortex matrix model is related to 2d gauge coupling
\begin{eqnarray}
\zeta=\frac{2\pi}{e^2}.
\end{eqnarray}
In order to decouple gravity from the gauge theories, the gauge coupling constant $g$ in vortex matrix model goes to infinity.
Then, the fields $\lambda$ and $D$ become the Lagrangian multipliers and produce the constraint;
\begin{eqnarray}{\label{Dterm}}
[B,B^{\dagger}]+ \sum_{i=1}^{N} I_i I^{\dagger }_i =\zeta \mathbb{I}_k.
\end{eqnarray}
Solutions of this equation characterize the moduli space for $k$-vortex 
\begin{eqnarray}
\mathcal{M}_{k, N}&=&\{(B, I_i) | [B,B^{\dagger}]+ \sum_{i=1}^{N} I_i I^{\dagger}_i =\zeta \mathbb{I}_k \}/ U(k). 
\end{eqnarray}
We can define the vortex partition function by using (\ref{knvortex})
\begin{eqnarray}
Z^{\mathcal{N}=(4,4)}=1+\sum_{k=1}^{\infty} Z^{\mathcal{N}=(4,4)}_{k, N} e^{2\pi (r+ i\theta) k}. 
\end{eqnarray}
 So far we have considered  only $N_c=N_f$ case, but vortex partition functions for general $N_c =N < N_f $
 is already given in \cite{HT1}, \cite{CT}. In general $N_f$ flavor case, 
the vortex partition functions are described by introducing additional $N_f-N_c$  dimensional reduced 2d chiral multiplets to zero dimension.
$(\tilde{I}_j, \tilde{\mu}_j), (j=1, \cdots, N_f-N\c)$.
The moduli space for $k$-vortex is modified to
\begin{eqnarray}
\mathcal{M}_{k, N_c, N_f}=\{(B, I_i, \tilde{I}_j) | [B,B^{\dagger}]+ \sum_{i=1}^{N_c} I_i I^{\dagger}_{i}
-\sum_{j=1}^{N_f-N_c}  \tilde{I}_j \tilde{I}^{\dagger}_j  =\zeta \I_{ k} \}/ U(k).  \nonumber \\
\end{eqnarray}
But, for simplicity, we restrict   our attention to $N_c  = N_f$ type vortices.

%%%%%%%%%%%%%%%%%%%%%%%%%%%%%%%%%%%%%%%%
\section{$\mathcal{N}=(2,2) $ vortex partition function}
In order to obtain $\mathcal{N}=(2,2) $ vortex partition functions,
we add a superpotential $\hat{W}(\Phi)$ to the 2d theory in (\ref{Lag44}).
\begin{eqnarray}
\int d^2 x d^2 \theta \mathrm{Tr} \hat{W}(\Phi).
\end{eqnarray}
This  breaks the $\mathcal{N}=(4,4)$ supersymmetry to $\mathcal{N}=(2,2)$ in two dimensions.
The superpotential contains  the quadratic term
\begin{eqnarray}\label{mass}
\hat{W}(\Phi)= M \Phi^2,
\end{eqnarray}
In the heavy mass limit, the bosonic part of Lagrangian in two dimensions is  obtained by setting $\Phi=0$ in  (\ref{Lag44}),
The vacuum in Higgs branch is still given by
\begin{eqnarray}
q^{a}_{i}=\sqrt{r} \delta^{a}_{i}, \quad \tilde{q}^{a}_{i}=0, \quad \sigma=0.
\end{eqnarray}
Apparently, the $\tilde{Q}$ fields are trivial, the vortex equation  is equivalently to vortex   for the following
  action  

\begin{eqnarray}\label{bos22f}
L_{boson}&=&- \Bigl[ \mathrm{Tr} \Bigl( \frac{1}{2e^2} F_{\mu \nu} F^{\mu \nu} +\frac{1}{2 e^2} D_{\mu} \sigma D^{\mu} \bar{\sigma}
+\frac{1}{2e^2} [\sigma, \bar{\sigma}]^2
  \nonumber \\
&&+\frac{e^2}{2} ( {q}_i  {q}^{\dagger}_i  -r \mathbb{I}_N)^2 + i \theta F_{12} \Bigr)
 +D_{\mu} {q}^{\dagger}_i D^{\mu} q_i  +  {q}^{\dagger}_i \{ \sigma, \bar{\sigma} \} {q}_i  
\Bigr]. \nonumber \\
\end{eqnarray}
This  is  bosonic parts of the Lagrangian for $\mathcal{N}=(2,2)$ $U(N)$ super Yang-Mills with $N$ fundamental 
chiral multiplets.

Let us consider  how the superpotential deformation (\ref{mass}) affects  the vortex matrix model (\ref{vortexmatrix}).
The vortices preserve the half of supersymmetry, so the  vortex matrix model is deformed  to preserve  half of the  $\mathcal{N}=(2,2)$ supersymmetry.
As discussed in \cite{ET}, in the presence of superpotential, vortex matrix model   is deformed by adding
\begin{eqnarray}
&&\mathrm{Tr} \int d \theta^+ \Xi \frac{\partial \hat{W}(\Sigma^{(0,2)})}{\partial \Sigma^{(0,2)}}|_{\bar{\theta}^+=0} + (\mathrm{h.c})
\nonumber \\  && \quad=G \frac{\partial \hat{W}(\phi) } {\partial \phi}+ \rho_- \frac{\partial^2 \hat{W}(\phi)}{\partial \phi^2} \bar{\lambda}_+
+(\mathrm{h.c}).
\end{eqnarray}
Here $\Sigma^{(0,2)}$ comes from the dimensional reduction of 2d $\mathcal{N}=(0,2)$ chiral multiplet,
\begin{eqnarray}
\Sigma^{(0,2)}=\phi-i \sqrt{2} \theta^+ \lambda_+ + \cdots.
\end{eqnarray}
$\Xi$ appears in the dimensional reduction of 2d $\mathcal{N}=(0,2)$ fermi multiplet,
\begin{eqnarray}
\Xi= \rho_-  - \sqrt{2}  \theta^+ G+ \cdots,
\end{eqnarray}
where $G$ is the auxiliary field in the fermi multiplet.

The left moving fermions   $\mu_-, {\mu}^{\dagger}_-, \rho_{-}$ and ${\rho}^{\dagger}_{-} $ are absent in the mass deformation.
Moreover in the heavy mass limit, $\Xi$ and $\Sigma^{(0,2)}$ multiplets are decoupled from the vortex matrix model. 
Thus the vortex partition functions for $\mathcal{N}=(2,2)$ theory consist of three pieces:
\begin{eqnarray}\label{vortex22}
{S'}_G&=&\mathrm{Tr} \Bigl( \frac{1}{2}[\varphi, \bar{\varphi}]^2+ D^2 +g^2 \zeta D +2\bar{\lambda}_{-}  [{\varphi}, \lambda_{-}]
  \Bigr), \nonumber \\
{S'}_{m}&=&\sum_{i=1}^{N} \Bigl( -I^{\dagger}_{ i} \{ \varphi, \bar{\varphi} \} I_{i}  
 +I^{\dagger}_{i} D I_i  \nonumber \\
&&-\sqrt{2} {\mu}^{\dagger}_{+ i } \bar{\varphi}  \mu_{+ i}%\bar{\mu}_{-} {\varphi}  \mu_{-}
+i \sqrt{2} [ I^{\dagger}_{ i} \lambda_{-} \mu_{+ i}  
-{\mu}^{\dagger}_{+ i} \bar{\lambda}_{-}  I_i ] \Bigr), \nonumber \\
{S'}_A&=&\mathrm{Tr} \Bigl( | [\varphi, B^{\dagger }] |^2  +D[B, B^{\dagger}]  \nonumber \\
&& -\sqrt{2} [\bar{\varphi},{\rho}^{\dagger}_+ ]  {\rho}_+
+  i \sqrt{2} \Bigl[ B[ {\rho}^{\dagger}_{+} , \bar{\lambda}_{-} ]+ B^{\dagger }  [ \lambda_{-},  \rho_{ +} ]  \Bigr] \Bigr). 
\end{eqnarray}
The $k$-vortex partition function for $\mathcal{N}=(2,2)$ super Yang-Mills theory is
\begin{eqnarray}\label{vortexpar22}
Z_{k,N} &=&\frac{1}{\mathrm{Vol} (U(k))} \int \mathcal{D} \varphi \mathcal{D} \bar{\varphi}
  \mathcal{D} B \mathcal{D} I \mathcal{D} \chi \mathcal{D} \eta \mathcal{D} \mu_+ \mathcal{D} \rho_+
     \exp (-\frac{1}{g^2} {S'}_G-{S'}_{m}-{S'}_{A}).\nonumber \\
\end{eqnarray}

%%%%%%%%%%%%%%%%%%%%%%%%%%%%%%%%
\section{Localization}
In this section, we compute the vortex partition function  in $\mathcal{N}=(2,2)$ super Yang-Mills theories by  equivariant localization formula for supersymmetric system  \cite{MNS1}, \cite{Nekrasov},  \cite{BFMT}.
The vortex partition function  given by (\ref{vortex22}) is invariant under the following supersymmetric transformation,
\begin{eqnarray}\label{equivariant}
&&Q_{\epsilon}\varphi=0, \nonumber \\
&&Q_{\epsilon} \bar{\varphi}=\eta, \quad Q_{\epsilon} \eta=[\varphi, \bar{\varphi}], \nonumber \\
&&Q_{\epsilon} D=[\varphi,\chi], \quad Q_{\epsilon} \chi=D, \nonumber \\
&&Q_{\epsilon} I_i=\mu_{+ i} , \quad Q_{\epsilon} \mu_{+ i}=\varphi I_i, \nonumber \\
&&Q_{\epsilon} I^{\dagger}_i=-{\mu}^{\dagger}_{+ i} , \quad Q_{\epsilon} {\mu}^{\dagger}_{+ i}=I^{\dagger}_i \varphi,   \nonumber \\
%&&Q \mu_{+ i}=0 \nonumber \\
%&&Q I^{\dagger}_i=-\bar{\mu}_{+ i} , \quad Q \bar{\mu}_{+ i}=-I^{\dagger}_i (\varphi -m_i)  \nonumber \\
%&&Q {\mu}^{\dagger}_{- i}=0 \nonumber \\
&&Q_{\epsilon} B=\rho_{+ }, \quad Q_{\epsilon} \rho_{+ }=[\varphi , B ] -\epsilon B, \nonumber \\ 
&&Q_{\epsilon} B^{\dagger}=-\rho^{\dagger}_{+} , \quad Q_{\epsilon} \rho^{\dagger}_{+ }=[\varphi , B^{\dagger} ] -\epsilon B^{\dagger}. 
\end{eqnarray}
Here we defined $\eta=-i({\lambda}_- +\bar{\lambda}_-)/\sqrt{2}$, $\chi=-i({\lambda}_- - \bar{\lambda}_-)/\sqrt{2}$ and rescaled
$\varphi \to -\sqrt{2} \varphi$.
The vortex partition function (\ref{vortexpar22}) can be written in $Q_{\epsilon}$-exact form.
For the abelian vortex case, we can apply the localization formula straightforwardly. 
But, $Q_{\epsilon}$ does not generate appropriate fixed points for the nonabelian case. We recall that vev of adjoint scalars in the vector multiplet play a crucial role to  localization formula work well and generate appropriate fixed points in Nekrasov partition functions in $\mathcal{N}=2$ super Yang-Mills theory. The twisted masses in chiral multiplets play a  role of vev of adjoint scalars  in the vortex partition functions at Higgs branch. 
We introduce the generic twisted masses $m_i, (i=1, \cdots, N)$, then the Lagrangian of 2d $\mathcal{N}=(2,2)$ super Yang-Mills
theory is modified as
\begin{eqnarray}\label{bos22}
L_{boson}&=&- \Bigl[ \mathrm{Tr} \Bigl( \frac{1}{2e^2} F_{\mu \nu} F^{\mu \nu} +\frac{1}{2 e^2} D_{\mu} \sigma D^{\mu} \bar{\sigma}
+\frac{1}{2e^2} [\sigma, \bar{\sigma}]^2 +\frac{e^2}{2} ( {q}  {q}^{\dagger}   -r \mathbb{I}_N)^2 \Bigr)
  \nonumber \\
&&+D_{\mu} {q}^{\dagger}_i D^{\mu} q_i +  {q}^{\dagger}_i \{ \sigma -m_i, \bar{\sigma} -m^*_i \} {q}_i   
\Bigr].
\end{eqnarray}
The vacuum is labeled by
\begin{eqnarray}
q^{a}_{i}=\sqrt{r} \delta^{a}_{i}, \quad \sigma^{a}_{b}= m^{a}_{i} \delta^{i}_{b}.
\end{eqnarray}
%This is the reason, we restrict $N_c=N_f$ case.
The introduction of twisted mass terms also introduce  twisted masses in vortex partitions.
The supersymmetry transformation is modified as 
\begin{eqnarray}
&& Q_{\epsilon} \mu_{+ i}=\varphi I_i \to  \quad Q_{\epsilon} \mu_{+ i}=(\varphi -m_i) I_i, \nonumber \\
&& \quad Q_{\epsilon} {\mu}^{\dagger}_{+ i}=I^{\dagger}_i \varphi \to  \quad Q_{\epsilon} {\mu}^{\dagger}_{+ i}=I^{\dagger}_i (-\varphi -m_i).
%&&Q \mu_{+ i}=0 \nonumber \\
\end{eqnarray}

The vortex partition function (\ref{vortex22}) can be written in $ Q_{\epsilon}$-exact form 
\begin{eqnarray}
{S'}_G&=& Q_{\epsilon} \mathrm{Tr} \left( \frac{1}{4} [\varphi, \bar{\varphi}] \eta + D \chi  +g^2 \zeta \chi \right), \nonumber \\
{S'}_{m}&=&\sum_{i=1}^{N_c} Q_{\epsilon} \Bigl[ {\mu}^{\dagger i}_{+}  (\bar{\varphi}-m^*_i)  I^{i }+{I}^{i \dagger} (\bar{\varphi}-m^*_i)   {\mu}^{i}_{+}+  I^{i \dagger} \chi I^{i} \Bigr], 
 \nonumber \\
{S'}_A&=& Q_{\epsilon} \mathrm{Tr} \Bigl[    -{B}^{ \dagger} [\bar{\varphi},   {\rho}_{+}]+{\rho}^{ \dagger}_{+} [\bar{\varphi}, {B}  ] +  \chi [B,B^{ \dagger}] \Bigr].
\end{eqnarray}
$Q_{\epsilon}$ is nilpotent up to   infinitesimal gauge transformation and with  flavor rotation.

Let us consider the localization method. First, we introduce the vector field $Q^*$ which acts on the fields  and generates the supersymmetry transformation

\begin{eqnarray}
Q^*&=&
[\varphi, \chi] \frac{\partial }{ \partial D }+ \eta \frac{\partial }{ \partial \bar{\varphi} } 
+ \mu_{+ i} \frac{\partial }{ \partial I_i } 
%+  {\mu}^{\dagger}_{+ i} \frac{\partial }{ \partial I^{\dagger}_i } 
+ \rho_{+} \frac{\partial }{ \partial B } %+  {\rho}^{\dagger}_{+ } \frac{\partial }{ \partial B^{\dagger}_i } 
\nonumber \\
&&+D \frac{\partial }{ \partial \chi}+[\varphi, \bar{\varphi}] \frac{\partial }{ \partial \eta}
+ (\varphi -m_i) I_i  \frac{\partial  }{ \partial \mu_+} 
%+I^{\dagger}_i (\varphi -m_i)   \frac{\partial  }{ \partial \mu^{\dagger}_+} \nonumber \\ &&
+ ([ {\varphi}, B ] -\epsilon B ) \frac{\partial }{ \partial {\rho}_+}
%+ ([ \bar{\varphi}, B^{\dagger} ] +\epsilon B^{\dagger} ) \frac{\partial }{ \partial {\rho}^{\dagger}_+} 
  \nonumber \\
&=&Q^{* i}_{\mathcal{F}}  \frac{\partial }{\partial \mathcal{B}_i }+ Q^{* i}_{\mathcal{B}}  \frac{\partial }{\partial \mathcal{F}_i }.
\end{eqnarray}
The critical points $Q^*=0$ is given by 
\begin{eqnarray}\label{fixedpoints}
&&(\varphi_I -m_i) I_{I i} =0, \nonumber \\
&& ({\varphi}_I - {\varphi}_J -\epsilon) B_{I J} =0.
\end{eqnarray}
All the other fields are zero. The $B$ and $I$ are all zero except for $B_{i, i-1}$ and $I_{1, 1}, I_{k_1+1, 2}, \cdots, I_{k_{N}+1, N}  $.
Here $k_{i}$'s are the partitions of $N$ integers  $( i=1, \cdots, N )$
 and satisfy a relation $k=\sum_{i} k_{i}$.

The localization formula \cite{BFMT} is expressed as contour integral
\begin{eqnarray}
Z_{k} =\oint \prod_{I=1}^{k} d \varphi_I \prod_{I \neq J} \frac{(\varphi_I - \varphi_J)}{\mathrm{Sdet} \mathcal{L}}.
\end{eqnarray}
The superdeterminant of $\mathcal{L}$ is defined by
\begin{eqnarray}
\mathrm{Sdet} \mathcal{L}=\mathrm{Sdet} \left(
 \matrix{    
    \frac{\partial Q^{* i}_{\mathcal{B}}}{\partial \mathcal{B}_j }  & \frac{\partial Q^{* i}_{\mathcal{B}}}{\partial \mathcal{F}_j }   
\cr \frac{\partial Q^{* i}_{\mathcal{F}}}{\partial \mathcal{B}_j }    &   \frac{\partial Q^{* i}_{\mathcal{F}}}{\partial \mathcal{F}_j }  }
\right),  
\end{eqnarray}
and 
\begin{eqnarray}
\mathrm{Sdet} 
\left(
 \matrix{    
    A  & B 
\cr C  & D    }
\right)  
&=&\mathrm{det} (A-BD^{-1} C) \mathrm{det} ({D})^{-1}. 
\end{eqnarray}

Thus, the vortex partition function $Z_k$ for $\mathcal{N}=(2,2)$ super Yang-Mills with twisted mass terms  becomes
\begin{eqnarray}\label{vorcon}
Z_{k} =\frac{1}{k! (2\pi i \epsilon)^k} \oint \prod_{I=1}^{k} d \varphi_I 
\Bigl( \prod_{I=1}^{k} \prod_{i=1}^{N} \frac{1}{\varphi_I -m_i} \Bigr) \prod_{I \neq J}
 \frac{\varphi_I - \varphi_J}{\varphi_I - \varphi_J - \epsilon }
\end{eqnarray}
For $N_c=N_f=1$ and $m_1=0$, the vortex partition function becomes

\begin{eqnarray}
Z_{k} =\frac{1}{k! (2\pi i \epsilon)^k} \oint \prod_{I=1}^{k} d \varphi_I \Bigl( \prod_{I=1}^{k} \frac{1}{\varphi_I } \Bigr) \prod_{I \neq J}
 \frac{\varphi_I - \varphi_J}{\varphi_I - \varphi_J - \epsilon }
\end{eqnarray}
This  reproduces the vortex partition function of abelian $k$-vortex in refined topological A-model amplitude of resolved conifold $\mathcal{O}(-1) \oplus \mathcal{O} (-1) \to \C P^1  $ in \cite{DGH}.
We do not   directly evaluate the contour integral of (\ref{vorcon}). 
Instead, in the next section, we calculate the equivariant character  which reproduce the residues of (\ref{vorcon})  \cite{NS}.

%%%%%%%%%%%%%%%%%%%%%%%%%%%%%%%%%%%%%%%%%%%%%%%%
\section{Vortex partition functions and equivariant character }
In this section, we calculate equivariant character of the vortex moduli spaces with twisted masses. 
This is similar to five  dimensional instanton counting or K-theoretic instanton counting in $\mathcal{N}=2$ super Yang-Mills theory.
We can recover the results in previous section by taking two dimensional limit.

First, we recall  the vortex moduli space $\mathcal{M}_{k, N}$ with $k-$vortex number is represented as \cite{EINOS}
\begin{eqnarray}{\label{Dterm}}
\mathcal{M}_{k, N}&=&\{(B, I) | [B,B^{\dagger}]+ I I^{\dagger} =\zeta \I_{k } \}/ U(k) \nonumber \\
&\simeq&\{(B_{\C}, I_{\C}) \}/ GL(k;\C).
\end{eqnarray}
where we  define the $k \times N$ matrix $I =(I_1, \cdots, I_{N})$.
We define the spaces  $V$ and $W$ on which $B_{\C}$ and $I_{\C}$ act, namely
$B_{\C} \in \mathrm{Hom}_{\C} (V, V)$, $I_{\C} \in \mathrm{Hom}_{\C} (W, V)$ with
 $\mathrm{dim}_{\C} V =k$ and $\mathrm{dim}_{\C} W =N$.

We define torus action $U_{\epsilon}(1) \times U(1)^{N}$ on $\mathcal{M}_{k, N}$.
First, we introduce  $U_{\epsilon}(1)$ action on $(B, I)$ by
\begin{eqnarray}\label{epsilonaction}
 U_{\epsilon}(1): (B_{\C}, I_{\C}) \to (q^{-1} B_{\C}, I_{\C}).
\end{eqnarray}
According to the action of $U_{\epsilon}(1)$, we modify the $B_{\C} \in \mathrm{Hom}_{\C} (V, V) \otimes Q$.
$Q$ is the one dimensional space on which $U_{\epsilon}(1)$ acts.
where $q=e^{-\beta \epsilon} \,(\epsilon \in \R )$.
Next, the action $ U(1)^{N} $ on the $(B_{\C}, I_{\C})$ is defined by
\begin{eqnarray}\label{naction}
U(1)^{N} : (B_{\C}, I_{\C}) \to ( B_{\C}, I_{\C} Q_{m}),
\end{eqnarray}
where $Q_m=\mathrm{diag} (e^{-\beta m_{1}}, e^{-\beta m_{2}}, \cdots,e^{-\beta m_{N}} ), \, Q_{m_{i}}=e^{-\beta m_i}$.

The $g \in \mathrm{Hom} ( U_{\epsilon} (1) \times U(1)^{N}, U (k))$ acts $(B, I)$ by
\begin{eqnarray}\label{kaction}
 g: ( B_{\C}, I_{\C}) \to (g(t) B_{\C} g(t)^{-1}, g(t) I_{\C}) \quad (t =(q, Q_{m}) \in U_{\epsilon} (1) \times U(1)^{N}). \nonumber \\ 
\end{eqnarray}
The fixed point conditions are written by (\ref{epsilonaction}), (\ref{naction}) and (\ref{kaction})
\begin{eqnarray}\label{fix}
(q^{-1} B_{\C}, I_{\C} Q_{m})= (g(t) B_{\C} g(t)^{-1}, g(t) I_{\C}).
\end{eqnarray}

We  set the $U(k)$ gauge transformation
\begin{eqnarray}
g(t)=\mathrm{diag}(e^{\beta \varphi_{1}}, e^{\beta \varphi_{2}}, \cdots, e^{\beta \varphi_{k}}).
\end{eqnarray}
Then, the fixed point conditions (\ref{fix}) become
\begin{eqnarray}
&&(\varphi_I-\varphi_J-\epsilon) B_{\C I J}=0, \nonumber \\
&&(\varphi_I-m_{i}) I_{\C I i}=0, \quad (I=1, \cdots, k \quad i=1, \cdots, N).
\end{eqnarray}
The solution of the fixed point conditions are
\begin{eqnarray}
\varphi_{I_{i}}=m_{i}+(I_{i}-1)\epsilon
\end{eqnarray}
which satisfies the fixed  points conditions (\ref{fixedpoints}).

We can decompose the representation space as follows
\begin{eqnarray}
&&W=\bigoplus_{i=1}^{N} W_{i}, \nonumber \\
&&V=\bigoplus_{i=1}^{N} \bigoplus_{n} V_{i}(n),
\end{eqnarray}
with
\begin{eqnarray}
&& W_{i}:=\{ w \in W | Q^{-1}_{m} w = Q^{-1}_{m_{i}} w\}, \nonumber \\
&& V_{i}(n):=\{ v \in V | g(t) v  = q^n  Q_{ m_{i}} v\}.
\end{eqnarray}
The dimensions of $ V_{i} (n) $ are  $0 \le  \mathrm{dim} V_{i} (n) \le 1$.
$(B_{\C},I_{\C})$ which satisfy the fixed point conditions (\ref{fix}) define the map
\begin{eqnarray}
&&B_{\C}:V_{i}(n)  \to V_{i}(n-1),  \nonumber \\
&&I_{\C}: W_{i}\to V_{i} (0). 
\end{eqnarray}
This means that the space $V$ is decomposed into $N$-tuple one dimensional partitions of $k_{i}$.
In the instanton case, recall  Nekrasov partition for $U(N)$ gauge theory is expressed by $N$-tuple two dimensional partitions; namely, Young diagrams. 

Next, we consider the tangent space of $\mathcal{M}_{k, N}$.
infinitesimal $\gl(k;\C)$ gauge transformation act,
$b: \gl (k;\C) \to \mathrm{Hom} (V, V \otimes Q) \oplus \mathrm{Hom} (W, V)   $
\begin{eqnarray}
b: \xi \to ([\xi, B], I \xi), \quad (\xi \in \gl(k;\C)).
\end{eqnarray}
Then, the tangent space $T_x {\mathcal{M}_{k,N}}$ of $\mathcal{M}_{k,N}$ is 
\begin{eqnarray}
T_x {\mathcal{M}_{k,N}}&=&\mathrm{Hom} (V, V \otimes Q) \oplus \mathrm{Hom} (W, V)/\mathrm{Im} b \nonumber \\
&=&\mathrm{Hom} (V, V \otimes Q) \oplus \mathrm{Hom} (W, V)/\mathrm{Hom} (V, V ).
\end{eqnarray}
We will use same   symbols for the characters and representation spaces. The characters are 
\begin{eqnarray}
V&=& \sum_{i=1}^{N} \sum_{l_{i}=1}^{k_{i}} q^{ l_{i}-1 } Q_{m_{i}}, \nonumber \\
W &=&\sum_{i=1}^{N}  Q_{m_{i}}, \nonumber \\
Q &=&  q^{-1}. 
\end{eqnarray}
where we defined $\mathrm{dim} \left( \oplus_{n} V_{i}(n) \right) =k_{i}$.
The character of the moduli space  becomes
\begin{eqnarray}
{ch}(T_x {\mathcal{M}_{k,N}})&=&(V^* \times   Q-V^* +W^*) \times V \nonumber \\
 &=&\sum_{i, \tilde{i}=1}^{N} Q_{{i \tilde{i}} } \sum_{l_{i}=1}^{k_{i}} 
q^{l_{i}- k_{\tilde{i}} -1}          \nonumber \\
 &=&\sum_{i, \tilde{i}=1}^{N} Q_{{i \tilde{i}} } \Bigl( \sum_{l_{i}=1}^{a_{i}} 
q^{l_{i}- k_{\tilde{i}} -1}  + \sum_{j_{i}=1}^{b_{i}} q^{j_{{i}}-1 }  \Bigr),       
\end{eqnarray}
where we have defined $Q_{{i \tilde{i}}}=Q_{m_{i}} Q^{-1}_{m_{\tilde{i}}}$ 
, $a_{i}=\mathrm{min}(k_{i}, k_{\tilde{i}})$ and $b_{i}=\mathrm{max}(k_{i}- k_{\tilde{i}},0)$. 
Finally, we obtain the K-theoretic vortex partition functions for $U(N)$ $\mathcal{N}=(2,2)$ super Yang-Mills theory
is
\begin{eqnarray}
Z^{\mathcal{N}=(2,2)}_{\mathrm{K-theoretic}}=\sum_{k=0}^{\infty} e^{2\pi (r+i \theta) k} \sum_{k_1+\cdots+k_N=k}  
\prod_{i, \tilde{i}=1}^{N}  \prod_{l_{i}=1}^{a_{i}} \prod_{j_{i}=1}^{b_{i}} 
%\nonumber \\  &&\qquad \times 
\frac{1 }{(1-Q_{ \tilde{i} i  } q^{ k_{\tilde{i}} +1 -l_{i}} ) (1-Q_{{\tilde{i} i } } q^{ 1-j_{i}  } )}. 
\nonumber \\
\end{eqnarray}
When we take the  two dimensional  limit,  $\beta \to 0$ with  $Q_{ m_{i}} =e^{- \beta m_{i} }$ 
and $q=e^{ -\beta \epsilon }$ and rescale $Z_{k, N}  $ by $\beta$.
Then, we obtain the residues of (\ref{vorcon})
\begin{eqnarray}
Z^{\mathcal{N}=(2,2)}=\sum_{k=0}^{\infty} e^{2\pi (r+i \theta) k} \sum_{k_1+\cdots+k_N=k}  \prod_{i, \tilde{i}=1}^{N}  
\prod_{l_{i}=1}^{a_{i}} \prod_{j_{\alpha}=1}^{b_{\alpha}} 
\frac{1 }{\bigl( ( k_{\tilde{i}} +1-l_{i} ) \epsilon - m_{i \tilde{i}}   \bigr) 
\bigl( ( 1-j_{i}) \epsilon - m_{i \tilde{i}}   \bigr)}, \nonumber \\
\end{eqnarray}
where we have defined $m_{i \tilde{i}}=m_{i }-m_{\tilde{i}}$.
For example, in abelian vortex case, the vortex partition function becomes
\begin{eqnarray}
Z^{\mathcal{N}=(2,2)}&=& \sum_{k=0}^{\infty} e^{2\pi (r+i \theta ) k}   \prod_{i=1}^{k } 
\frac{1 }{( k +1-i) \epsilon } 
 \nonumber \\
&=& \exp \Bigl( \frac{ e^{2\pi (r+i \theta ) }}{\epsilon } \Bigr).
\end{eqnarray}

\section{Discussion}
We have calculated the vortex partition  of  $\mathcal{N}=(2,2)$ $U(N)$ super Yang-Mills theory with $N$ fundamental chiral 
multiplets. The introduction of twisted mass terms break the non-abelian symmetry to collection of $U(1)^N$ symmetry.
Thus, our results is the collection of $N$-tuple abelian vortex rather than purely non-abelian vortex. Moreover, the localization with twisted mass terms works well in the case of $N_c=N_f$. How the vortex partitions become in the general flavor cases $N_c  < N_f$ ? When the gauge group is $U(1)$, \cite{DGH} obtain general $N_f$ flavor vortex partition functions by refined BPS state counting and they also reveal  the vortex partition function for $U(1)$ gauge group  with $N_f$ fundamental chiral multiplets is related to  $J$-function of $\C P^{N_f-1}$; the  generating function of genus zero Gromov-Witten invariants  for $\C P^{N_f-1}$ up to appropriate parameters identification. 
At strong coupling limit $e^2 \to \infty$, the vortex moduli space for $U(N_c)$ super Yang-Mills  with $N_f$ flavor 
becomes  the moduli space of  holomorphic maps from the complex plane $\C$ to the 
grassmann manifold $\mathrm{Gr}(N_c,N_f)$ \cite{EINOS}.
Rouhgly speaking, we can regard tht the vortex partitions functions is  the geometric index(or volume) of the moduli space 
of holomorphic maps. So we expect to the purely non-abelian vortex partition functions for
$U(N_c)$  gauge group with $N_f$ fundamental flavor is related to the  $J$-function  for
the grassmann manifold $\mathrm{Gr}(N_c,N_f)$,  \cite{HV}, \cite{LLLY}, \cite{BFK}.

\section*{Acknowledgments}
We are grateful to K. Ito, T. Kimura and K. Otha   for valuable discussions.
We are also grateful to K. Sugiyama for  reading of the manuscript and  helpful comments.

\begingroup\raggedright

\end{document}